# An ER-Model-Based Framework for Case Notion Selection in Object-Centric Processes

Akhil Kumar

Smeal College of Business, Penn State University
University Park, PA 16802, USA
Akhilkumar@psu.edu

**Abstract.** Object-centric process mining operates on event logs where each event references multiple objects of different types. A fundamental challenge is defining a *case notion* — the grouping of events into coherent process execution instances — without which process discovery and conformance checking cannot proceed. Existing approaches either flatten the log to a single object type (losing inter-object coordination) or use the connected component of the object graph (creating overly complex cases due to resource-like objects). We propose an Entity-Relationship-schema-guided framework that identifies the primary entity (PE) type anchoring each process execution, classifies other entity types as secondary coordination entities or resources, and defines cases as connected components of the primary and secondary entity object relationship graph. The resulting case notion is shown to produce strict partitions with automatic transitive closure. The framework is illustrated with an order management process and validated on a 1,000-event OCEL log producing 40 structurally coherent cases.

## 1 Introduction

Process mining extracts process knowledge from event logs recorded by information systems [1,3]. Traditional process mining assumes each event belongs to exactly one *case* — a process execution instance identified by a case identifier. This assumption breaks down in modern enterprise systems where a single activity involves multiple business objects simultaneously: a *create package* event in an order management system references multiple orders and one package; a *conduct interview* event references an applicant, a vacancy, and a recruiter. This results in 1-to-many and many-to-many relationships in the log that should inform the analysis of such logs [16].

Object-Centric Event Logs (OCEL) [14] address this by recording all participating objects for each event. However, the fundamental problem of defining a case notion remains open. Without cases, the event log cannot be fed to standard process discovery algorithms, conformance checking cannot be performed, and performance indicators cannot be computed per execution instance.

A natural baseline is to define one case per connected component of the full object graph. This correctly handles chained object relationships but produces spurious cases when resource objects — products, trucks, recruiters — appear across multiple independent executions and act as bridges, merging unrelated process instances. Van Detten et al. [11] propose an approach based on transitive closure but it operates at the activity level and does not guarantee a strict partition.

We propose an ER-schema-guided approach that grounds the case notion in the business data model rather than in log statistics alone. Our contributions are: (1) formal definitions of primary, resource, secondary and parent entity scope derived from ER cardinalities; (2) a case notion based on connected components of the object relationship graph over primary and secondary entity objects with strict partition and transitive closure; and (3) a two-level process discovery framework separating PE lifecycle discovery from synchronization structure discovery.

The next section offers a background and discusses related work. Then Section 3 motivates our work with an example of order management and derives an ER diagram for it. Next, Section 4 describes our framework in detail. Later, Section 5 discusses how this case notion leads to a two-level process discovery approach that can use traditional process mining algorithms. Finally, Section 6 gives a discussion and Section 7 concludes the paper.

## 2 Background and Related Work

**Object-Centric Process Discovery**. The object-centric directly-follows graph (OC-DFG) [6] extends the classical directly-follows graph to multiple object types. Van der Aalst [3] proposes object-centric Petri nets (OCPN) as a formal model with synchronization arcs and provides a discovery algorithm based on inductive mining per object type; van der Aalst and Berti [5] establish the foundational OCPN discovery framework. Van der Aalst [4] extends this to object-centric conformance checking. Our two-level discovery framework builds on the OCPN formalism but precedes OCPN discovery with explicit case notion selection grounded in the ER schema. In [2], van der Aalst et al. propose extending data models with a behavioral perspective, modelling cardinality constraints over activities — complementary to our structural approach. The approach in [10] uses Multiple Viewpoint (MVP) models that relate events through objects and activities through classes. MVP models allow for the extraction of classical event logs using different viewpoints so that existing process mining techniques can be used for each viewpoint.

**Case Notion in Object-Centric Processes**. Defining a case notion in OCPM has received increasing attention and is the core topic of this paper. Esser and Fahland [12] show that relational structure is essential for representing and querying multidimensional process data — our framework extends this by using the ER schema to guide case formation directly. A recent manifesto [18] highlights the importance of an adequate case notion. Adams et al. [8] define cases and variants for object-centric event data by working at the activity-type level using connected components of the directly-follows graph. Their approach is data-driven and does not distinguish a resource from coordination entities, so resource objects can still merge independent executions into one case. Our framework resolves this through explicit entity classification grounded in the ER schema.

The advanced case notion in [11] is based on transitive relationships while stopping at diverging object types, thus partially preventing spurious merging. This schema-free approach does not guarantee strict partitions and does not handle joint-execution cases where multiple PE instances coordinate through a shared bridge entity. Our framework guarantees strict partitions by construction and correctly handles joint execution using primary and secondary entities.

Fahland and Montali [13] explore what constitutes an object-centric case conceptually, identifying several candidate definitions at different granularities. Their work is definitional and complementary to ours: where they explore the conceptual space, we provide an operational framework with formal guarantees.

Khayatbashi et al. [15] propose "process scopes" as an alternative case notion, using activity-level annotations to define case boundaries on BPIC logs transformed to OCEL format. Their approach is data-driven and schema-free; ours requires an ER schema but benefits from a deeper understanding of the underlying data model and its implications.

De Weerdt et al. [20] study case notion selection in process mining benchmarking, showing that case notion choice substantially affects discovered model quality - motivating the need for principled selection criteria such as ours.

Table 1 summarizes the comparison with other approaches on the case notion.

**Table 1**. Comparison of case notion approaches.

| **Approach** | **Schema required** | **Strict partition** | **No spurious merging** | **Joint execution** |
|---|---|---|---|---|
| Van Detten et al. [11] | No | No | Partial | No |
| Adams et al. [8] | No | Yes | No | Yes |
| Khayatbashi et al. [15] | No | No | No | No |
| Fahland & Montali [13] | Conceptual | — | — | — |
| This paper | Yes (ER) | Yes | Yes | Yes |

## 3 Motivating Example: Order Management Process

### 3.1 The Process

We motivate the framework with an order management process. Customers place orders; each order may contain multiple items. Items are picked from the warehouse, packed into packages, and shipped to customers. A single package may contain items from multiple orders placed by the same customer, enabling consolidated shipment. Table 2 shows eight representative events from the log.

**Table 2**. Eight representative events from the order management log.

| e# | Activity | Timestamp | Orders | Pkgs | Items | Customer |
|---|---|---|---|---|---|---|
| **e1** | place order | 2019-05-20 09:07 | 990001 | — | 880001–03 | MP |
| **e2** | place order | 2019-05-20 10:35 | 990002 | — | 880005–08 | GP |
| **e3** | pick item | 2019-05-20 10:38 | 990002 | — | 880006 | GP |
| **e8** | item out of stock | 2019-05-20 13:54 | 990001 | — | 880003 | MP |
| **e32** | confirm order | 2019-05-21 12:19 | 990002 | — | 880006, 880007 | GP |
| **e91** | pay order | 2019-05-22 14:47 | 990002 | — | 880005–08 | GP |
| **e121** | create package | 2019-05-23 11:00 | 990002,990011 | 660001 | 880006,32 … | GP |
| **e206** | pkg delivered | 2019-05-24 15:54 | 990001,990025 | 660002 … | 880002,03 | MP |

Event e121 (create package) is particularly significant: it references two orders (990002 and 990011) and one package (660001) simultaneously. This is a *synchronization event* — a coordination point where multiple process execution instances converge on a shared artifact. Event e206 (package delivered) similarly references two orders (990001 and 990025) via package 660002. These multi-order events motivate the need for a case notion that can handle coordination across order instances when some activities (e.g. create package) are shared across multiple orders.

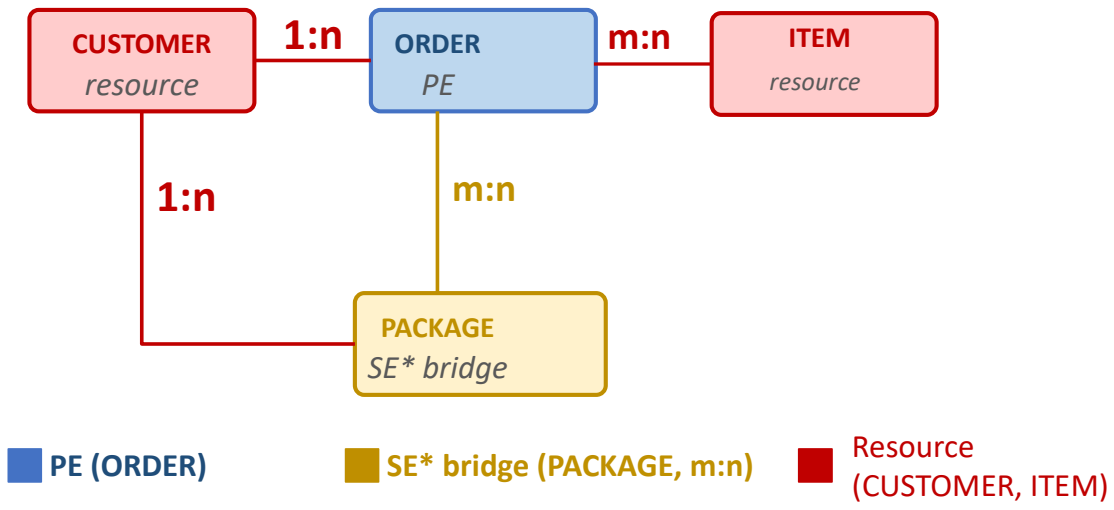

**Fig. 1**. ER schema for the order management process

### 3.2 The ER Diagram

An ER Schema S is a tuple (ET, R, card) where ET represents entity types drawn from a business domain; R is set of relationships among the entities; and card : R → {1:1, 1:n, n:1, m:n} assigns cardinality to each relationship. Sometimes cardinalities are included in the representation of R itself and S is denoted as (ET,R). Fig. 1 shows the ER schema for this process consisting of entity types, relationships and their cardinalities (1:1, 1:n, m:n). Four entity types are present: CUSTOMER, ORDER, ITEM, and PACKAGE. Their relationships are: CUSTOMER $\rightarrow_{1:n}$ ORDER; ORDER $\rightarrow_{m:n}$ ITEM (catalog product); ORDER $\rightarrow_{m:n}$ PACKAGE (consolidated shipment). ITEM is a **Resource entity (RE)**, i.e., items are catalog products shared across orders from different customers. CUSTOMER is a **parent** entity (ET_par) in that one customer owns many independent orders. PACKAGE is a **Secondary (bridge) entity (SE)**. Each entity has suitable attributes that are omitted here.

## 4 Formal Framework

### 4.1 Object-Centric Event Log

***Definition 1 (Object-Centric Event Log).*** Let OT be a set of object types and OI a set of object identifiers. An *object-centric event log* is a finite set $L = \{e_1,\ldots,e_n\}$ where each event e = (e.id, e.activity, e.time, e.objects, e.attr) and e.objects ⊆ OI × OT is a finite set of typed object references such that each identifier appears with at most one type.
The following are derived from L:
**O** = ⋃_{e∈L} {o : (o,t) ∈ e.objects} — the set of all object instances in L
**type**: O → OT where type(o) = t for the unique t such that (o,t) ∈ e.objects for some e
**objects(e)** = {o : (o,t) ∈ e.objects} — the set of object ids referenced by event e
**O2O** = $\{(o_1,o_2) : \exists e \in L, o_1,o_2 \in objects(e), o_1 \neq o_2\}$ — set of object co-occurrence pairs

*Example 1.* Event e121: e.objects = {(990002, ORDER), (990011, ORDER), (660001, PACKAGE), (880006, ITEM), (880032, ITEM)}. Derived: objects(e121) = {990002, 990011, 660001, 880006, 880032}. O2O edges from e121 include (990002, 660001), (990011, 660001), (990002, 990011) — the last is a same-type ORDER–ORDER pair arising from co-occurrence in a shared packaging event.

### 4.2 Entity Classification

***Definition 2 (Primary Entity — PE).*** Consider a Schema S = (ET, R), where ET is the set of entity types and R is the set of relationships (with associated cardinalities) among the entities. PE is the entity type t ∈ ET that represents the primary business object of the process, as identified from the ER schema S and domain knowledge.

*Example 2.* In the order management process the primary business object is the Order — it is what customers place, what warehouse staff fulfil, and what is tracked from placement to delivery. Thus, PE = ORDER.

The choice of PE can be empirically corroborated from the log using two statistics. For entity type t, define:

**cov(t)** = fraction of events in L referencing ≥ 1 instance of type t,
**share(t)** = fraction of events in L referencing > 1 instance of type t.

A well-chosen PE should have *high coverage* (*cov*)— it participates in most process events — and *low share* — its instances tend to appear independently rather than co-occurring in the same event. Default thresholds $\theta_{cov}$ = 0.5 and $\theta_{share}$ = 0.3 provide a practical filter: types failing either threshold are unlikely to be viable PE candidates regardless of schema structure. In the sample log of Table 1, ORDER has cov = 1.0 and share = 0.25, consistent with its role as PE. PACKAGE has cov = 0.125 — it appears in only a minority of events, confirming it is a secondary entity rather than the process backbone. In general, a PE candidate should have cov(t) substantially above $\theta_{cov}$. Note that $\theta_{share}$ and $\theta_{cov}$ are empirical tunable defaults.

Before the next definition, we note that *some entity types can be excluded* from case formation directly from the ER schema, without reference to log statistics. Specifically, any entity type t for which card(t → PE) = 1:n in S — meaning each PE instance has exactly one parent instance of type t, and acts as a classifier or owner of PE rather than a coordination artifact. Since each PE instance carries exactly one value of t (analogous to an attribute), t adds no structural information to the O2O graph. In the order management schema, CUSTOMER →1:n→ ORDER means each order belongs to exactly one customer. Thus, CUSTOMER is treated as an attribute-like parent of ORDER and excluded from case formation. We denote the set of such **Parent entities as ET_par** = {t ∈ ET\{PE} : card(t → PE) = 1:n in S}.

***Definition 3 (Resource Entity).*** ET_res = {t ∈ ET \ {PE} : share(t) ≥ θ_share}, where θ_share is a user-defined threshold.

Resource entities are catalog or physical objects whose instances are shared across multiple events — they are referenced by the process rather than created or managed by it. Resource instances are excluded from the O2O graph.

*Example 3.* From Table 1: share(ITEM) = 6/8 = 0.750 $\geq$ 0.3 $\rightarrow$ ITEM $\in$ ET_res. share(PACKAGE) = 0/8 = 0.000 < 0.3 $\rightarrow$ PACKAGE $\notin$ ET_res. ET_res = {ITEM}.

***Definition 4 (Secondary Entity— SE).*** An entity type t $\in$ ET \ (ET_res $\cup$ ET_par $\cup$ {PE}) is a *secondary entity* if card(PE $\rightarrow$ t) $\in$ {1:n, m:n}. The set of all secondary entity types is: SE* = {t $\in$ ET \ (ET_res $\cup$ ET_par $\cup$ {PE}) : card(PE $\rightarrow$ t) $\in$ {1:n, m:n}}

*Example 4.* ET \(ET_res $\cup$ ET_par $\cup$ {PE}) = {PACKAGE}.
card(ORDER$\rightarrow$PACKAGE) = m:n $\in$ {1:n, m:n} $\rightarrow$ PACKAGE $\in$ SE*.
SE* = {PACKAGE}.

### 4.3 Case Notion

The case notion partitions events into cases by computing connected components of the object relationship graph over PE and SE* instances.

***Definition 5 (Case Notion).***
Let L be an OCEL log, PE the primary entity type, SE* the set of secondary entity types, ET_res the resource entity set, and ET_par the parent entity set.
**Step 1 — Object relationship graph:** G(L) = (V, E) where
V = {o $\in$ O : type(o) $\notin$ ET_res $\cup$ ET_par}
E = {($o_1$, $o_2$) $\in$ O2O : $o_1$, $o_2$ $\in$ V}
**Step 2 — Connected components:** {$K_1$,...,$K_m$} = connected_components(G(L)) where each $K_k \subseteq$ V satisfies:
(i) **Connectivity:** for any $o_1$, $o_2 \in K_k$ there exists a path ($o_1$, $u_1$,..., $o_2$) in G(L)
(ii) **Maximality:** there is no o $\in$ V \ $K_k$ connected by a path in G(L) to any o' $\in K_k$
(iii) **Partition:** $K_1$,...,$K_m$ are pairwise disjoint and $K_1 \cup ... \cup K_m$ = V
**Step 3 — Case event sets:** Let O_PE = {o $\in$ O : type(o) = PE}. Define events($K_k$) = {e $\in$ L : objects(e) $\cap K_k \cap$ O_PE $\neq \emptyset$}. By construction of E, for any event e all objects in objects(e) $\cap$ V lie within the same component, so events($K_1$),...,events($K_m$) are pairwise disjoint and cover all events referencing at least one PE object.
**Step 4 — Case collection:**
cn*(L) = {$K_k \cap$ O_PE : k = 1,...,m, $K_k \cap$ O_PE $\neq \emptyset$}
Each case is the set of PE instances in one connected component. SE* objects serve as bridges in G(L) to connect PE instances but are not part of the case output.

*Example 5.* (see log of Table 1)
<u>Step 1</u>: {(990002, 660001), (990011, 660001), (990002, 990011), (990001, 660002), (990025, 660002), (990001, 990025)}.
<u>Step 2</u>: $K_1$ = {990002, 990011, 660001} — connected via 660001, maximal; $K_2$ = {990001, 990025, 660002} — connected via 660002, maximal; $K_1 \cap K_2 = \emptyset$, $K_1 \cup K_2$ = V.
<u>Step 3</u>: events($K_1$) = {e2, e3, e32, e91, e121}; events($K_2$) = {e1, e8, e206} — disjoint, all 8 events covered. <u>Step 4</u>: cn*(L) = {{990002, 990011}, {990001, 990025}}.

***Definition 6* (Case Identity and Identifier).**
Let $C_k = K_k \cap O_PE$ be a case from Definition 5. The case identifier is:
$cid(C_k) = MD5(sort(C_k))$
where $sort(C_k)$ concatenates the PE instance ids of $C_k$ in ascending lexicographic order separated by hyphens, and MD5 is applied to the UTF-8 encoding of the resulting string.
A case with $|C_k| = 1$ is a *single-executor* case — one PE instance with its own independent lifecycle. A case with $|C_k| > 1$ is a *joint-execution* case — multiple PE instances whose lifecycles synchronize through at least one shared SE* object instance.

*Example 6.* $C_1 = \{990002, 990011\} \rightarrow |C_1| = 2 \rightarrow$ joint-execution: orders 990002 and 990011 synchronize through package 660001. $cid(C_1)$ = MD5("990002-990011") = "cbf7…". $C_2 = \{990001, 990025\} \rightarrow cid(C_2)$ = MD5("990001-990025") = "19a77…".

*Remark 1*. We can verify that all 8 events of Table 1 are assigned, with no overlap and no omissions. The two cases form a strict partition. Event e121 belongs to $C_1$ because both 990002 and 990011 (and 660001) are in $C_1$. Event e206 belongs to $C_2$ for the same reason.

*Remark 2*. Three properties of this identifier are that it is:

- **Deterministic** — same PE membership always produces the same hash
- **Fixed length** — regardless of whether the case has 1 or 11 orders
- **Collision-resistant** — with a standard hash (SHA-256, MD5) collisions are negligible in practice

The one limitation worth acknowledging: the hash is opaque — you cannot read off the orders from the identifier. But $C_k$ is always stored alongside $cid(C_k)$ and serves as the human-readable description of the case..

### 4.4 Properties

The following properties (P1-P3) hold for cn*(L) by construction of Definition 5.
**P1 (Strict Partition).** Every PE instance in O_PE appears in exactly one case in cn*(L), and $events(K_1),...,events(K_m)$ are pairwise disjoint and cover all events referencing at least one PE object.
**P2 (Transitive Closure).** If two PE instances are connected through any chain of O2O edges over non-resource, non-parent objects they belong to the same case.
**P3 (PE Anchor).** Every case in cn*(L) contains at least one PE instance. A component with no PE objects is excluded from cn*(L) by the condition $K_k \cap O_PE \neq \emptyset$ in Step 4.
Each property is directly readable from the definition steps that guarantee it. P1 from Step 2(iii) and Step 3; P2 from Step 2(i); and P3 from Step 4.

### 4.5 Evaluation on a 1,000-Event Log

We apply the framework to a 1,000-event OCEL log (accessible from http://akhilkumar2.github.io/home/order_log2_1K) with 114 orders, 50 packages, and 17 customers, extracted from the order management process available at ocel-standard.org. PE = ORDER, ET_res = {ITEM}, ET_par = {CUSTOMER}, SE* = {PACKAGE}. Applying Definition 5 yields 40 cases with Overlap = 0 and Omit = 0 - a strict partition of all 1,000 events. Table 3 shows a representative selection.

**Table 3**. Representative cases from the 1,000-event log. MD5 = hash function

| Case id | Customer | #Orders | #Packages | #Events | Type |
|---|---|---|---|---|---|
| MD5(990002-… -990086) | GP | 11 | 4 | 110 | Joint |
| MD5(990006-…-990085) | CR | 9 | 4 | 91 | Joint |
| MD5(990001-…-990098) | MP | 6 | 4 | 68 | Joint |
| MD5(990004-…-990094) | JG | 8 | 3 | 63 | Joint |
| MD5(990051-…-990089) | LM | 4 | 2 | 35 | Joint |
| … | … | … | … | … | … |

The largest case belongs to GP — 11 orders transitively connected through four packages (660001, 660013, 660027, 660039), each bridging two or more orders, giving a case id cid = MD5("990002-990009-990011-990012-990013-990036-990046-990057-990070-990074-990086"). No single package links all 11 orders — it is the transitive closure of Definition 5 that correctly groups them into one joint-execution case. CR's case has 9 orders and 4 packages (91 events), and MP's has 6 orders and 4 packages (68 events). All 17 customers have at least one joint-execution case — no customer has all single-executor cases — confirming that consolidated shipment is the norm in this process. Of the 40 cases, 21 are complete (at least one PACKAGE linked) and 19 are incomplete (no package observed at the log boundary). Incomplete cases are valid cases awaiting their SE* events and do not violate the partition guarantee.

Fig. 2 shows a partial O2O graph with edges between orders and packages. The dashed lines illustrate how a connected component occurs when multiple orders are included in one package.

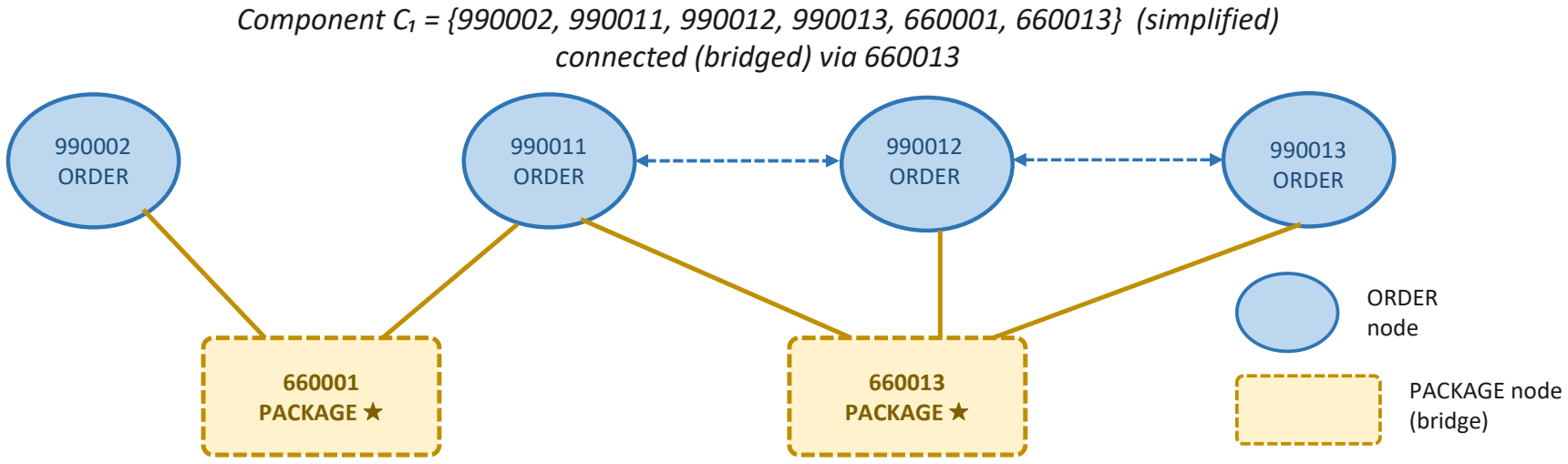

**Fig. 2**. An example to illustrate bridging orders through shared package

## 5 Two-Level Process Discovery: What the Case Notion Enables

The case notion defined in Section 4 is the essential prerequisite for process discovery on object-centric event logs. Without a principled partition of the log into cases, standard discovery algorithms cannot be applied: there is no well-defined unit of execution to project, no trace to mine, and no basis for measuring model quality. This section applies the two-level structure to the 1,000-event order management log.

**Level 1 — Order Lifecycle**. Projecting one trace per PE (ORDER) instance yields 114 order traces and 109 distinct variants, reflecting variation in pick item repetitions (ranging from 0 to 10 per order), out-of-stock occurrences and interleaving artefacts arising from projecting joint-execution cases onto individual order traces. The backbone lifecycle is: place order → confirm order → (pick item)* → pay order, with an out-of-stock exception loop (item out of stock → reorder item → retry pick item). Further, events referencing both ORDER and PACKAGE objects appear in the per-order traces as coordination markers connecting Level 1 to Level 2. The exception activities — item out of stock (96 events) and failed delivery (9 events) — are directly visible in the per-order traces without requiring case-level aggregation.

**Table 4**. Package lifecycle trace variants (50 packages, 1K log).

| # | Trace variant | Count |
|---|---|---|
| 1 | create package → send package → package delivered | 40 |
| 2 | create package → send package → failed delivery → package delivered | 5 |
| 3 | create package → send package (incomplete) | 2 |
| 4 | create package → send package → failed delivery (incomplete) | 1 |
| 5 | create package → send package → failed delivery → failed delivery → failed delivery → package delivered | 1 |
| 6 | create package (incomplete) | 1 |

**Level 2 — Package Lifecycle**. Projecting one trace per PACKAGE instance yields 50 package traces with only 6 distinct variants, shown in Table 4. The dominant variant (create package → send package → package delivered) accounts for 40 of 50 packages. The remaining 10 show incomplete executions or failed delivery retry loops. This low variant count — compared to 109 at Level 1 — confirms the package lifecycle is structurally simpler than the order lifecycle, exactly as the ER schema predicts. PACKAGE is a coordination artifact (SE bridge) with a fixed three-step lifecycle, while ORDER is the primary entity driving full process variation.

**Sync transitions**. Events referring both ORDER and PACKAGE objects are the sync transitions connecting the two levels: create package (50 events), send package (49), package delivered (46), and failed delivery (9). All four carry ORDER–PACKAGE cardinality m:n. The arc multiplicity k — the number of ORDER tokens firing together in a single sync event — ranges from 1 to 6 in the 1K log. This is distinct from case size: GP's case has 11 orders, yet no single sync event involves all 11 simultaneously. Instead, the 11 orders are distributed across four separate packaging events (k=5, k=6, k=4, k=4 orders), and it is the transitive closure through shared packages that groups all 11 into one case. The two sub-logs — 114 order traces at Level 1 and 50 package traces at Level 2 — can each be fed directly to any standard discovery algorithm such as the inductive miner [17].

The framework supports |SE*| + 1 discovery levels — one per SE* type plus the PE level. Deeper recursive nesting, where a SE* type itself has sub-entities requiring further decomposition, is a natural, further extension of this work. In the order management process SE* = {PACKAGE}, yielding two levels; a logistics process with SE* = {PACKAGE, SHIPMENT} would yield three. This multi-level structure has a natural counterpart in process execution: Steinau et al. [19] propose coordination processes to manage interactions among related process instances at run-time, showing that 1:n and m:n inter-process relationships give rise to distinct coordination layers. Our framework recovers those layers analytically from the event log — the Level 2 sub-log

reveals the same coordination structure that Steinau et al. [19] design explicitly at execution time.

## 6 Discussion

Rather than taking a purely data-centric approach, we develop a new case notion that is grounded in the Entity-Relationship model. Each design decision (which entity is primary, resource, secondary, parent) corresponds to a business semantic claim that can be validated against domain knowledge. By defining cases as connected components, the framework guarantees meaningful, non-overlapping partitions by construction.

Our framework's primary contribution is a case notion selection method. By grounding PE selection in an ER schema structure rather than log statistics alone, and by excluding resource entities from the object graph before computing connected components, the framework avoids two failure modes of prior approaches: losing inter-object coordination (flattening) and merging independent executions through shared resources (over-connection). The strict partition, transitive closure, and "PE anchor" guarantees follow directly from the connected component construction with no optimization required. Other approaches fail to guarantee strict partitioning.

The key design decisions each have a direct semantic justification. Resource exclusion (ET_res) prevents entities that appear across independent executions — ITEM — from bridging unrelated cases. The 1:1 and n:1 exclusion from SE* prevents attribute extensions of PE from inflating structural complexity. The case identifier obtained from hashing the order numbers of the orders in a case serves as a fixed-length key that is stable.

The case partitions also enable a clean decomposition into two sub-logs with strikingly different characteristics: 114 order traces (109 variants) at Level 1 and 50 package traces (6 variants) at Level 2. This asymmetry — the PE lifecycle is complex and variable, the SE* lifecycle is simple and regular — is a direct consequence of the ER schema's role in entity classification and is invisible in the raw log without the partition. A natural question after constructing the case partition is: how do we know the entity classification is correct? We cannot label cases as right or wrong without ground truth. Instead, the ER schema itself provides a reference: if PACKAGE is correctly classified as SE*, then the schema's m:n cardinality between ORDER and PACKAGE should accurately predict the coordination structure observed in each case — cases with more orders should have proportionally more package links. Structural complexity (SC) is measured as the actual coordination load of each case instance by the O2O edges observed in the log. Structural homogeneity H measures how consistently the actual SC matches the baseline prediction across all complete cases. A high H confirms the classification is coherent: the schema predicts what the log shows. A low H signals misclassification — if ITEM were incorrectly included in SE*, item-heavy cases would have disproportionately high SC relative to the schema's prediction and H would fall sharply. These concepts should be developed in subsequent work.

In terms of limitations, the framework requires an ER schema as input that must be created correctly by an expert with domain knowledge. The $\theta_{share}$ and $\theta_{cov}$ must be suitably tuned. Moreover, our testing has been limited to a single log. Further testing on more complex logs with multiple SE's would be helpful in future work. Another limitation is that only one SE is considered in this work. Further extension, say, by adding INVOICE (ORDER→1:n→INVOICE) and PAYMENT (INVOICE→n:1→PAYMENT) to the schema of Fig. 1 illustrates the framework's boundary. INVOICE qualifies as SE* through its direct 1:n relationship with ORDER and produces a Level 2 invoice lifecycle. *PAYMENT has no* ***direct*** *relationship with ORDER in the ER schema — it is reachable from ORDER only through INVOICE. Definition 4 considers only direct PE→t relationships, so PAYMENT does not qualify for SE*. It is excluded from V entirely and cannot create spurious bridges. Its lifecycle is not discoverable as a Level 2 sub-log in the current framework. Whether PAYMENT should be included by extending Definition 4 to cover transitively reachable entity types depends on domain semantics — if bulk payments cover multiple orders they behave like resources (share(PAYMENT) would be high), while per-order payments would be valid secondary entities. Depending upon the business practice of a supplier, INVOICE can also be related to PACKAGE if invoices are generated by package and not order. These extensions deserve further investigation.

## 7 Conclusion

We have presented an ER-schema-guided framework for case notion selection in object-centric process mining. The framework classifies entity types as primary (PE), secondary (SE), resource (RE), or parent (ET_par) using schema cardinality and the share statistic; constructs the object relationship graph over primary and secondary instances only; and defines cases as connected components of that graph. The result is a strict partitioning of the event log with automatic transitive closure — guaranteed by construction, with no optimization or post-processing. This is the basis of the proposed case notion.

Each case is assigned a fixed length Case Identifier to serve as a stable surrogate key compatible with standard ERP id assignment. The case notion allows for a two-level process discovery framework. The case partition decomposes the log into two sub-logs ready for standard discovery algorithms: a level 1 log that captures the full PE lifecycle complexity, and a level 2 log that captures the (structurally simpler) SE lifecycle. The contrast between the two levels is interesting — one that is invisible in the raw log and becomes apparent only upon applying the ER-grounded case notion.

Two quality measures are proposed to ground the framework empirically. Structural complexity SC quantifies the coordination load of each case, while structural homogeneity H measures how consistently actual case structures match ER schema predictions. The preliminary ideas on these measures should be developed further.

Future work includes adaptive threshold selection for resource entity classification, and empirical evaluation across the OCEL benchmark suite. The two-level framework also needs to be implemented and tested extensively. Finally, treatment of multiple SE's deserves more exploration.